\documentclass[preprint,prd,aps,showpacs,preprintnumbers,amsmath,amssymb]{revtex4-1}
\usepackage{graphics,epsfig,subfigure}
\usepackage{diagbox}
\usepackage[usenames]{color}
\usepackage{color}
\usepackage{graphicx}
\usepackage{amsfonts}
\usepackage{indentfirst}
\usepackage{booktabs}
\usepackage{hyperref}
\hypersetup{hypertex=ture,backref=true,colorlinks=ture,linkcolor=blue,anchorcolor=blue,citecolor=blue}
\usepackage{float}
\usepackage{multirow}
\usepackage[section]{placeins}
\usepackage{array}
\usepackage{amsmath}

\begin{document}
	\title{\Large \bf Off-shell Hessian thermodynamic  stability of higher-curvature black holes}
	\author{Chen-Hao Hao$^{1,2}$, Jieci Wang$^{1,2}$}
	\email{jcwang@hunnu.edu.cn, corresponding author}
	\affiliation{$^{1}$Department of Physics, Key Laboratory of Low Dimensional Quantum Structures and Quantum  Control of Ministry of Education, and Synergetic Innovation Center for Quantum Effects and Applications, Hunan Normal
		University, Changsha, Hunan 410081, P. R. China\\
		$^{2}$ Institute of Interdisciplinary Studies, Hunan Normal University, Changsha, Hunan 410081, P. R. China}
	
	\begin{abstract}
		We develop a branch-sensitive thermodynamic framework for higher-curvature
		black holes using the off-shell Gibbs free energy \(G_{\rm off}\)
		and the Wald entropy \(S_W\) as the basic data. On fixed-parameter slices,
		equilibrium black holes are stationary points of \(G_{\rm off}\), and their
		local stability is governed by the Hessian \(H=S'_W(r_h)T'(r_h)\), rather
		than by the temperature slope alone. For the five-dimensional charged regular
		AdS black hole in quasi-topological gravity, \(S_W\) remains monotonic on the
		physical branch, so the usual temperature-slope rule is recovered only as a
		special consequence. The same off-shell structure also gives the local
		\(A_3\) cusp normal form near criticality, yielding the mean-field \(1/2\)
		branch separation exponent and explaining why smooth nondegenerate observables,
		such as the Lyapunov exponent, inherit the same scaling. In
		Lovelock black holes, \(S'_W\) can change sign on
		non-planar branches, reversing the temperature slope stability assignment. However, on ghost-free and branch-regular Lovelock exteriors \(S'_W\) remains
		positive.
		Thus the off-shell Hessian criterion also diagnoses why the ordinary slope rule
		is protected on physically admissible black holes branches.
	\end{abstract}

	\maketitle
	
	\section{Introduction}
	\label{sec:introduction}
	Black hole thermodynamics provides a central arena for exploring the interplay
	between gravitation, statistical physics, and quantum field theory \cite{Witten:2024upt,Padmanabhan:2009vy,Bardeen:1973gs,Bekenstein:1973ur,Iyer:1994ys,Strominger:1997eq,Wei:2015iwa,Maldacena:1996gb,Wei:2019uqg}. In asymptotically anti-de Sitter spacetime in particular, black holes exhibit a rich phase structure, owing both to their intrinsic thermodynamic properties and to the deep significance of the AdS/CFT correspondence \cite{Maldacena:1997re,Aharony:1999ti}. The Hawking-Page transition of the Schwarzschild-AdS
	black hole describes a transition between thermal AdS and a large black hole
	phase \cite{Hawking:1982dh,Witten:1998zw}, while charged AdS black holes exhibit first-order small/large
	black hole transitions closely analogous to the liquid-gas transition of
	ordinary fluids \cite{Chamblin:1999hg,Kastor:2009wy,Kubiznak:2012wp}. This analogy becomes particularly sharp in extended
	phase space thermodynamics, where the cosmological constant is interpreted as
	thermodynamic pressure and the black hole mass as enthalpy \cite{Kastor:2009wy,Dolan:2011xt}. These developments have shifted the emphasis from black holes as isolated
	solutions of gravitational field equations to black holes as thermodynamic
	systems whose temperature, entropy, and free energy encode the structure of
	the underlying solution space \cite{Cai:2013qga,Cheng:2016bpx,Xu:2022jyp,Xu:2024iji,Altamirano:2013uqa,Altamirano:2013ane,Frassino:2014pha,Hennigar:2016xwd,Bai:2023woh,Dolan:2014vba,Hu:2024ldp,Ahmed:2022kyv,Yang:2025xck,Chen:2023pgs,Wu:2024asq}.
	
	Several complementary schemes have recently been used to classify black hole phase structure. One is a local geometric classification based on the number of nondegenerate extrema of the temperature function \(T(r_h)\) \cite{Zhang:2025inl}. A second treats black hole states as thermodynamic defects carrying winding numbers \cite{Wei:2024gfz,Liu:2025iyl,Wei:2026upy}. A third uses complexified thermodynamic functions, generalized free energies, and the zero structure on the associated Riemann surface \cite{Xu:2023vyj,Xu:2025jrk}. In Einstein gravity these viewpoints are closely related, at least on the real thermodynamic domain: the extrema of \(T(r_h)\) determine the number of branches, their stability ordering, the local topological charges, and the corresponding complex zero-counting picture \cite{Zhang:2026sht}. This suggests that the different classifications are projections of the same branch structure. It also raises the question addressed here: what remains of this dictionary beyond Einstein gravity, where the entropy need not obey the area law and the sign of \(T'(r_h)\) no longer determines local stability by itself?

	This question is physically motivated because higher-curvature terms are expected whenever general relativity is treated as an effective theory. Such corrections arise in string theory, M theory, and low-energy effective field theory \cite{Zwiebach:1985uq,Gross:1986iv,Green:1997as,Donoghue:1994dn}, and provide a controlled way to probe ultraviolet and strong-curvature effects in black hole physics, such as Lovelock gravity \cite{Lovelock:1971yv,Boulware:1985wk,Cai:2001dz}. More recently, quasi-topological gravity has attracted considerable interest. Its higher-order curvature terms preserve tractable equations in the static sector while encoding corrections beyond the Einstein-Hilbert action \cite{Bueno:2024dgm,Bueno:2024zsx,Oliva:2010eb,Myers:2010ru,Dehghani:2011vu,Ahmed:2017jod,Cisterna:2017umf,Frolov:2024hhe,Frolov:2025ddw,DiFilippo:2024mwm,Aguayo:2025xfi,Tsuda:2026xjc,Borissova:2026dlz,Ling:2025ncw,Hao:2025utc,Hao:2026pwl,Li:2026mam,PinedoSoto:2026hfm,Saridakis:2026oyf}. The charged regular AdS black hole in quasi-topological gravity is therefore a natural testbed. Its thermodynamics is explicit, including the Wald entropy, equation of state, Gibbs free energy, coexistence curve, and branch-dependent Lyapunov behavior \cite{Wang:2024zlq,Hennigar:2025yqm,Xie:2025auj}. Meanwhile, it differs from the Einstein case: the entropy is Wald entropy rather than area entropy, the higher-curvature coupling can enter the first law, and the physical horizon branch is restricted. Thus the Einstein-gravity dictionary cannot simply be copied. It must be reformulated in terms of the appropriate thermodynamic potential, stability criterion, and complex domain.
	
	In this paper, we focus on the local thermodynamic stability criterion for
	higher-curvature black holes. Using the off-shell Gibbs free energy
	\(G_{\rm off}\), we show that on fixed-parameter branches the correct
	Hessian criterion is \(H=S'_W(r_h)T'(r_h)\), rather than the temperature slope
	alone. We first test this criterion in the five-dimensional charged regular
	AdS black hole in quasi-topological gravity; consequently, the
	usual temperature slope rule is recovered only as a special case. We then
	contrast this with Lovelock black holes, for which \(S'_W\propto
	\Upsilon'(g_h)\) can change sign on non-planar branches. A formal negative entropy orientation would
	reverse the temperature-slope stability assignment, but in ghost-free and
	branch-regular Lovelock exteriors this possibility is excluded. Thus the
	entropy orientation is not only part of the stability criterion, but also a
	diagnostic of branch admissibility.
	
	The paper is organized as follows. In Section~\ref{sec:offshell_general}, we develop
	the general off-shell Gibbs free energy construction and the Hessian
	stability criterion. In Section~\ref{sec:qt_black_hole}, we
	apply the framework to the five-dimensional charged regular AdS black hole in
	quasi-topological gravity, analyze the
	off-shell free energy profile, and extract the local cusp structure near the
	critical point. In Section~\ref{sec:lovelock_black_holes}, we extend the Hessian criterion
	to Lovelock black holes and show how a change in the entropy orientation can
	reverse the usual stability assignment. Section~\ref{sec:conclusions}
	summarizes the results and discusses possible extensions. Finally, we list more examples of high-curvature black holes in the Appendix~\ref{app} to illustrate the generality of our framework.
	
	\section{Off-shell thermodynamics in higher-curvature gravity}
	\label{sec:offshell_general}
	
	In this section we formulate the local thermodynamic stability criterion. The framework is independent of any particular black hole solution. The only required ingredients are a fixed-parameter first law, a well defined entropy
	functional, and a one-dimensional physical branch parametrized by the horizon radius \(r_h\).
	This formulation is especially useful in higher-curvature gravity, where the black hole entropy
	is the Wald entropy rather than the area entropy, and the usual temperature slope rule need
	not be a fundamental stability criterion.
	
	\subsection{Fixed-parameter first law and off-shell Gibbs free energy}
	\label{subsec:general_offshell}
	
	We consider a family of static black holes described by a horizon radius \(r_h\). The mass,
	temperature, and entropy are regarded as functions of \(r_h\) and of a set of external
	thermodynamic parameters. These parameters may include the thermodynamic pressure \(P\), conserved
	charges \(Q\), and higher-curvature couplings collectively denoted by \(\lambda_a\).
	In an extended thermodynamic description, the first law takes the schematic form
	\begin{equation}
		dM
		=
		T\,dS_W
		+ \Phi\,dQ
		+ V\,dP
		+ \sum_a \Psi_a\,d\lambda_a ,
		\label{eq:general_first_law_full}
	\end{equation}
	where \(S_W\) is the Wald entropy, $\Phi$ is its conjugate potential to conserved charge \(Q\), and \(\Psi_a\) are the thermodynamic potentials conjugate to the higher-curvature couplings.
	
	We focus on local stability on fixed-parameter slices. That is, when the horizon
	radius is varied, the quantities $(P, Q, \lambda_a)$
	are held fixed. On such a slice the first law reduces to
	\begin{equation}
		dM = T\,dS_W .
		\label{eq:fixed_parameter_first_law}
	\end{equation}
	Equivalently, we have
	\begin{equation}
		\frac{dM}{dr_h}
		=
		T(r_h)\frac{dS_W}{dr_h}.
		\label{eq:mass_entropy_relation}
	\end{equation}
	This identity is the only thermodynamic input needed below.
	
	We now introduce an external heat bath of temperature \(T_0\). The off-shell Gibbs free
	energy is defined by \cite{York:1986it,Wei:2022dzw}
	\begin{equation}
		G_{\rm off}(r_h;T_0)
		=
		M(r_h)-T_0 S_W(r_h),
		\label{eq:general_goff}
	\end{equation}
	where all external parameters other than \(T_0\) are kept fixed. The adjective ``off-shell''
	means that \(r_h\) is not required to satisfy the equilibrium condition \(T(r_h)=T_0\).
	Thus $G_{\rm off}$ assigns a thermodynamic potential to each horizon configuration on the chosen branch.
	
	Differentiating \eqref{eq:general_goff} and using \eqref{eq:mass_entropy_relation}, one obtains
	\begin{equation}
		\frac{dG_{\rm off}}{dr_h}
		=
		\frac{dM}{dr_h}
		-
		T_0\frac{dS_W}{dr_h}
		=
		\bigl[T(r_h)-T_0\bigr]S_W'(r_h),
		\label{eq:general_goff_derivative}
	\end{equation}
	where a prime denotes differentiation with respect to \(r_h\). Therefore, whenever
	$S_W'(r_h)\neq 0$,
	the stationary condition
	\begin{equation}
		\frac{dG_{\rm off}}{dr_h}=0
		\label{eq:goff_stationary_condition}
	\end{equation}
	is equivalent to the thermal equilibrium condition
	$T(r_h)=T_0$.
	Thus the usual temperature equation is recovered as the stationarity condition of \(G_{\rm off}\), rather than being imposed as a separate variational principle. The fundamental object is the off-shell free energy constructed from the appropriate Wald entropy.
	
	Equation \eqref{eq:general_goff_derivative} also shows that points with
	$S_W'(r_h)=0$
	require special care. At such points, $dG_{\rm off}/dr_h$ may vanish even when the configuration is not in thermal equilibrium with the bath. Hence an entropy turning
	point is not automatically an equilibrium black hole. Rather, it signals a degeneracy of the
	entropy coordinate on the chosen thermodynamic branch. In what follows, the one-dimensional local stability analysis below will be restricted to branch segments on which \(S'_W(r_h)\neq0\).
	
	\subsection{General Hessian stability criterion}
	\label{subsec:general_hessian}
	
	The local stability of an equilibrium branch is determined by the second variation of
	\(G_{\rm off}\). Differentiating \eqref{eq:general_goff_derivative} once more, we obtain 
	\begin{equation}
		\frac{d^2G_{\rm off}}{dr_h^2}
		=
		T'(r_h)S_W'(r_h)
		+
		\bigl[T(r_h)-T_0\bigr]S_W''(r_h).
		\label{eq:general_goff_second_derivative}
	\end{equation}
	At an equilibrium point \(r_h=r_i\), where \(T(r_i)=T_0\), the second term vanishes.
	Therefore the local Hessian is
	\begin{equation}
		H_i
		\equiv
		\left.
		\frac{d^2G_{\rm off}}{dr_h^2}
		\right|_{r_h=r_i}
		=
		\left.
		S_W'(r_h)T'(r_h)
		\right|_{r_h=r_i}.
		\label{eq:general_hessian}
	\end{equation}
	The equilibrium point is locally stable against fluctuations of the horizon radius on the
	fixed-parameter slice if
	$H_i>0$,
	and locally unstable if
	$H_i<0$.
	The marginal case \(H_i=0\) is not decided by the quadratic variation and
	requires the higher-order expansion of \(G_{\rm off}\). This occurs, for
	example, at temperature turning points \(T'(r_i)=0\) or entropy-degenerate
	points \(S'_W(r_i)=0\).
	This is the basic local stability criterion used in this paper. It differs conceptually from
	the familiar temperature slope rule. In Einstein gravity with the area law, one can find that 
	\begin{equation}
		S_W(r_h)\propto r_h^{D-2},
		\qquad
		S_W'(r_h)>0
		\quad
		(r_h>0),
		\label{eq:einstein_entropy_positive}
	\end{equation}
	and therefore
	\begin{equation}
		{\rm sign}\,H_i
		=
		{\rm sign}\,T'(r_i).
		\label{eq:einstein_slope_rule}
	\end{equation}
	In that special case, a positive temperature slope indicates local stability and a negative
	temperature slope indicates local instability.
	
	In higher-curvature gravity, however, the entropy is generally not the area entropy. The
	sign of \(S_W'(r_h)\) is model dependent and branch dependent. Consequently, the sign of
	\(T'(r_h)\) alone does not determine local stability. The correct criterion is
	\begin{equation}
		{\rm sign}\,H_i
		=
		{\rm sign}\!\left[S_W'(r_i)T'(r_i)\right].
		\label{eq:general_sign_rule}
	\end{equation}
	If \(S_W'(r_i)>0\), the usual temperature slope rule is recovered. If \(S_W'(r_i)<0\), the
	temperature slope assignment is reversed.
	
	Thus a branch with positive temperature slope can be locally unstable, while a branch with
	negative temperature slope can be locally stable. This possibility has no analogue in
	Einstein gravity with a positively oriented area entropy.
	
	The same conclusion follows from the heat capacity. On a fixed-parameter slice, one obtains 
	\begin{equation}
		C
		=
		T\left(\frac{\partial S_W}{\partial T}\right)
		=
		T\frac{S_W'(r_h)}{T'(r_h)}.
		\label{eq:general_heat_capacity}
	\end{equation}
	For positive temperature,
	\begin{equation}
		{\rm sign}\,C
		=
		{\rm sign}\!\left[\frac{S_W'(r_h)}{T'(r_h)}\right].
		\label{eq:heat_capacity_sign}
	\end{equation}
	
	Away from temperature turning points, the heat-capacity criterion and the off-shell Hessian criterion give the same sign for $T>0$. However, this equivalence of signs does not make the Hessian formulation a mere rewriting of the heat capacity. The heat capacity is a response function involving the ratio $S'_W/T'$, whereas the Hessian follows from the second variation of the off-shell free energy and directly identifies local minima and maxima of the thermodynamic potential. It also avoids dividing by $T'(r_h)$ and, more importantly, separates the equilibrium equation $T(r_h)=T_0$ from the entropy orientation $S'_W(r_h)$. This separation is the key point: the usual temperature-slope rule is valid only when the Wald entropy is positively oriented along the branch.
	
	\subsection{Entropy orientation on the thermodynamic branch}
	\label{subsec:entropy_orientation}
	
	The factor \(S_W'(r_h)\) has a simple interpretation. It measures the orientation of the
	entropy coordinate along the one-dimensional configuration space parametrized by \(r_h\). When
	$S_W'(r_h)>0$,
	the entropy increases with the horizon radius, and the branch has the same orientation as
	in Einstein gravity. In this case the local minima and maxima of \(G_{\rm off}\) are ordered
	in the same way as the increasing and decreasing segments of \(T(r_h)\).
	When
	$S_W'(r_h)<0$,
	the entropy decreases as the horizon radius increases. The entropy coordinate then has
	the opposite orientation to the radius coordinate. The off-shell Hessian detects this reversal
	through the factor \(S_W'\). As a result, the usual temperature slope rule is reversed even
	though the equilibrium equation remains \(T(r_h)=T_0\).
	
	This distinction is central in higher-curvature thermodynamics. The function \(T(r_h)\)
	determines where equilibrium branches occur, but \(T(r_h)\) alone does not determine
	whether those branches are stable. Stability also depends on the entropy measure on the
	branch. The off-shell free energy separates these two pieces of information:
	\begin{equation}
		\frac{dG_{\rm off}}{dr_h}
		=
		\underbrace{\bigl[T(r_h)-T_0\bigr]}_{\text{equilibrium equation}}
		\underbrace{S_W'(r_h)}_{\text{entropy orientation}} .
		\label{eq:equilibrium_orientation_split}
	\end{equation}
	
	This perspective also changes the interpretation of the ordinary temperature slope rule. When the rule remains valid in a higher-curvature theory, such as quasi-topological gravity, this is not a purely kinematical fact following from the shape of $T(r_h)$ alone. Rather, it means that the physical branch carries a positive entropy orientation, $S'_W(r_h)>0$, throughout the relevant domain. In this sense the validity of the usual slope rule becomes a nontrivial statement about the admissibility of the thermodynamic branch. As will be seen explicitly in Lovelock gravity, ghost-free and branch-regular exteriors can protect the positive sign of $S'_W$, thereby protecting the Einstein-like slope assignment. Conversely, any violation of this entropy orientation would signal either a genuine reversal of the local stability ordering or a pathology of the underlying branch.
	
	\section{Quasi-topological regular AdS black hole}
	\label{sec:qt_black_hole}
	
	We now illustrate the general stability criterion developed in Section~\ref{sec:offshell_general}
	using the five-dimensional charged regular AdS black hole in
	quasi-topological gravity.
	This example
	is useful because the entropy is a genuine Wald entropy, rather than the area entropy, but
	it remains monotonic on the physical branch. Therefore the ordinary temperature slope
	rule is recovered only after the entropy orientation has been checked explicitly.
	
	\subsection{Action, solution and physical branch}
	\label{subsec:qt_action_solution}
	
	We consider quasi-topological gravity coupled to the Maxwell field,
	\begin{equation}
		I=I_{\rm QT}+I_{\rm EM},
		\label{eq:qt_total_action}
	\end{equation}
	where
	\begin{equation}
		I_{\rm QT}
		=
		\frac{1}{16\pi G}
		\int d^Dx\sqrt{|g|}
		\left[
		R-2\Lambda+\sum_{n=2}^{\infty}\tilde{\alpha}_n Z_n
		\right],
		\qquad
		I_{\rm EM}
		=
		-\frac{1}{16\pi G}
		\int d^Dx\sqrt{|g|}\,
		F_{\mu\nu}F^{\mu\nu}.
		\label{eq:qt_action}
	\end{equation}
	Here \(R\) is the Ricci scalar, \(\Lambda\) is the cosmological constant, \(F_{\mu\nu}\) is the Maxwell field strength, \(Z_n\) denotes the quasi-topological density of order \(n\), and \(\tilde\alpha_n\) is the corresponding higher-curvature coupling. The explicit contractions entering \(Z_n\) will not be needed below and can be found in \cite{Bueno:2024dgm}. The role of these quasi-topological densities is to encode higher-curvature corrections while preserving a tractable reduced field equation in the static, spherically symmetric sector. Moreover, it has been conjectured that all higher-order curvature gravities can
	be characterized as generalized quasi-topological gravity theories via field redefinitions \cite{Bueno:2019ltp}. In this sense, the model provides a useful effective theory laboratory for testing how black hole thermodynamics is modified in higher-curvature gravity.
	
	We take the metric ansatz
	\begin{equation}
		ds^2
		=
		-N(r)^2f(r)dt^2
		+
		\frac{dr^2}{f(r)}
		+
		r^2d\Omega_{D-2}^2 .
		\label{eq:qt_metric_ansatz}
	\end{equation}
	The reduced field equations imply that \(N(r)\) is a constant. We fix the time normalization so that \(N=1\), consistently with the
	thermodynamic quantities used below.
	
	We focus on the branch generated by the special choice of higher-curvature couplings
	\begin{equation}
		\tilde{\alpha}_n
		=
		\frac{
			\bigl[1-(-1)^n\bigr](D-2)\Gamma(n/2)
		}{
			2\sqrt{\pi}(D-2n)\Gamma((n+1)/2)
		}
		\alpha^{\,n-1}.
		\label{eq:qt_special_couplings}
	\end{equation}
	This choice is not unique, but it leads to a regular black hole geometry in the
	static sector \cite{Hennigar:2025yqm,PinedoSoto:2026hfm,Hao:2025utc}. The metric function can be written as
	\begin{equation}
		f(r)
		=
		1-
		\frac{r^2\mathcal{S}(r)}
		{\sqrt{1+\alpha^2\mathcal{S}(r)^2}},
		\label{eq:qt_metric_function}
	\end{equation}
	with
	\begin{equation}
		\mathcal{S}(r)
		=
		\frac{2\Lambda}{(D-1)(D-2)}
		+
		\frac{m}{r^{D-1}}
		-
		\frac{q^2}{r^{2D-4}},
		\label{eq:qt_S_function}
	\end{equation}
	where
	\begin{equation}
		m
		=
		\frac{16\pi GM}{(D-2)\Omega_{D-2}},
		\qquad
		q
		=
		\frac{8\pi GQ}
		{\sqrt{2(D-2)(D-3)}\,\Omega_{D-2}} .
		\label{eq:qt_mq_definitions}
	\end{equation}
	
	The horizon radius is determined by \(f(r_h)=0\). Hence
	\begin{equation}
		\frac{r_h^2\mathcal{S}(r_h)}
		{\sqrt{1+\alpha^2\mathcal{S}(r_h)^2}}
		=
		1.
		\label{eq:qt_horizon_condition}
	\end{equation}
	For the branch considered here, we take
	$\mathcal{S}(r_h)>0$.
	Squaring \eqref{eq:qt_horizon_condition}, one obtains 
	\begin{equation}
		\bigl(r_h^4-\alpha^2\bigr)\mathcal{S}(r_h)^2=1.
		\label{eq:qt_squared_horizon_condition}
	\end{equation}
	Therefore a real horizon on this branch requires
	\begin{equation}
		r_h^4>\alpha^2,
		\label{eq:qt_horizon_bound_general}
	\end{equation}
	which becomes
	\begin{equation}
		r_h>\sqrt{\alpha}
		\label{eq:qt_horizon_bound}
	\end{equation}
	for \(\alpha>0\).
	This lower bound is the endpoint of the analytic horizon branch, not an ordinary
	thermodynamic equilibrium state. The positive temperature canonical sector may be a
	proper subset of the interval \(r_h>\sqrt{\alpha}\).
	
	The spacetime is regular at the center. As \(r\to0\), the metric function behaves as
	\begin{equation}
		f(r)
		=
		1+\frac{r^2}{|\alpha|}
		+
		O(r^3),
		\label{eq:qt_regular_core}
	\end{equation}
	and for nonzero charge, the \(q^2/r^{2D-4}\) term dominates near the origin, leading to curvature invariants remaining finite. In what follows we specialize to \(D=5\), for which \(\Omega_3=2\pi^2\).
	
	\subsection{Extended thermodynamics}
	\label{subsec:qt_thermodynamics}
	
	In the extended phase space, the cosmological constant is interpreted as pressure,
	\begin{equation}
		P
		=
		-\frac{\Lambda}{8\pi G}
		=
		\frac{(D-1)(D-2)}{16\pi G l^2}.
		\label{eq:qt_pressure_general}
	\end{equation}
	
	The black hole mass is interpreted as enthalpy. The extended first law takes the form
	\begin{equation}
		dM
		=
		T\,dS_W
		+
		\Phi_{\rm EM}\,dQ
		+
		V\,dP
		+
		\Psi\,d\alpha .
		\label{eq:qt_first_law_full}
	\end{equation}
	Although \(\alpha\) appears as a thermodynamic variable in the full extended first law, the
	local analysis below is performed on fixed-\((P,Q,\alpha)\) slices. Thus the relevant
	one-dimensional first law is
	\begin{equation}
		dM=T\,dS_W .
		\label{eq:qt_fixed_first_law}
	\end{equation}
	
	For the five-dimensional charged regular AdS black hole, the thermodynamic quantities are \cite{Hennigar:2025yqm,Borissova:2026rbi}
	\begin{equation}
		M
		=
		\frac{3\pi}{8G}
		r_h^4
		\left(
		\frac{1}{\sqrt{r_h^4-\alpha^2}}
		+
		\frac{1}{l^2}
		+
		\frac{q^2}{r_h^6}
		\right),
		\label{eq:qt_mass}
	\end{equation}
	\begin{equation}
		S_W
		=
		\frac{\pi^2r_h^3}{2G}\,
		{}_2F_1
		\left(
		\frac{3}{2},
		-\frac{3}{4};
		\frac{1}{4};
		\frac{\alpha^2}{r_h^4}
		\right),
		\label{eq:qt_wald_entropy}
	\end{equation}
	\begin{equation}
		Q
		=
		\frac{\sqrt{3}\pi}{2G}q,
		\qquad
		\Phi_{\rm EM}
		=
		\frac{\sqrt{3}}{2}\frac{q}{r_h^2},
		\label{eq:qt_charge_potential}
	\end{equation}
	and
	\begin{equation}
		V
		=
		\frac{\pi^2}{2}r_h^4 .
		\label{eq:qt_volume}
	\end{equation}
	The crucial difference from Einstein gravity is already visible in \eqref{eq:qt_wald_entropy}.
	The entropy is not the area entropy; it contains a hypergeometric function and must be
	used as the correct entropy measure on the thermodynamic branch.
	
\begin{figure}[htbp]
		\centering
		\includegraphics[width=0.8\linewidth]{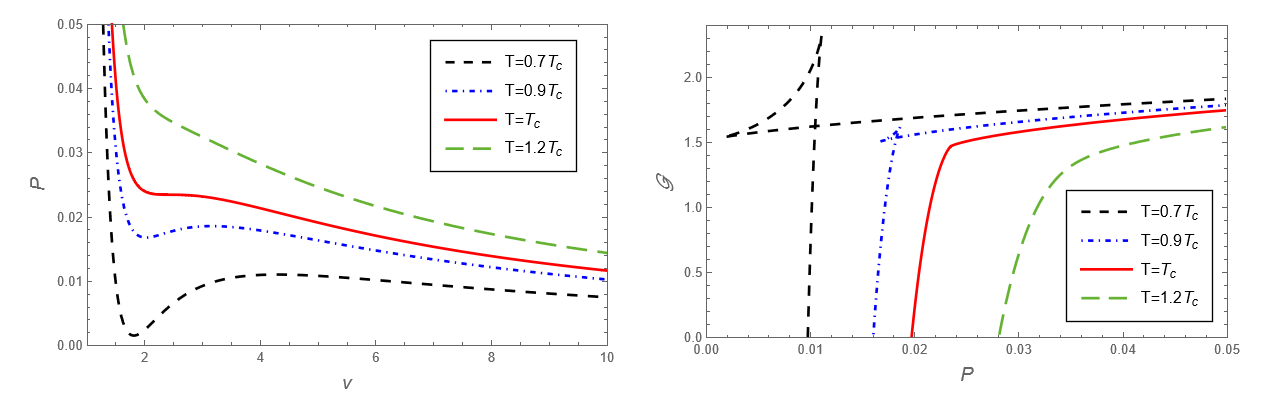}
		\caption{Parameters are chosen as $q=2/3$, $b=4/5$. For $T<T_c$, the $P$--$v$ isotherm develops two spinodal points, which bound the unstable intermediate black hole branch. The two outer segments correspond to locally stable small and large black holes, while the middle segment is locally unstable. In the $\mathcal{G}$--$P$ plane, these two spinodal points map to the two turning points of the swallowtail. The swallowtail intersection marks the first-order small/large black hole transition, whereas the turning points themselves mark the limits of metastability. At $T=T_c$, the two spinodal points merge into a critical inflection point and the swallowtail disappears.}
		\label{p1}
	\end{figure}

	It is useful to introduce the specific volume \(v\) and the parameter \(b\), the equation of state is defined on the physical domain \(v>b\), equivalently
	\(r_h>\sqrt{\alpha}\).
	\begin{equation}
		v=\frac{4G}{3}r_h,
		\qquad
		b=\frac{4G}{3}\sqrt{\alpha}.
		\label{eq:qt_v_b_definitions}
	\end{equation}
	Setting \(G=1\), the equation of state becomes
	\begin{equation}
		P
		=
		\frac{T v^5}{(v^4-b^4)^{3/2}}
		-
		\frac{2(v^4-2b^4)}{3\pi (v^4-b^4)^{3/2}}
		+
		\frac{512q^2}{243\pi v^6}.
		\label{eq:qt_equation_of_state}
	\end{equation}
	Equivalently, we obtain
	\begin{equation}
		T(v;P,q,b)
		=
		\frac{(v^4-b^4)^{3/2}}{v^5}
		\left[
		P
		+
		\frac{2(v^4-2b^4)}{3\pi (v^4-b^4)^{3/2}}
		-
		\frac{512q^2}{243\pi v^6}
		\right],
		\label{eq:qt_temperature_v}
	\end{equation}
	which can be written as
	\begin{equation}
		T(r_h;P,q,\alpha)
		=
		\frac{4P}{3}
		\frac{(r_h^4-\alpha^2)^{3/2}}{r_h^5}
		+
		\frac{r_h^4-2\alpha^2}{2\pi r_h^5}
		-
		\frac{q^2(r_h^4-\alpha^2)^{3/2}}{2\pi r_h^{11}},
		\label{eq:qt_temperature_rh}
	\end{equation}
	in terms of \(r_h\).
	The on-shell Gibbs free energy is
	\begin{equation}
		G_{\rm on}=M-TS_W,
		\label{eq:qt_on_shell_gibbs_def}
	\end{equation}
	where \(T\) is the Hawking temperature. For the parametric \(G\)--\(P\) curve at fixed \(T\), we use the equation of state to eliminate \(P\) in favor of \((r_h,T,q,\alpha)\). This gives
	\begin{align}
		G_{\rm on}(r_h;T)
		=
		&
		\frac{9\pi q^2}{16r_h^2}
		+
		\frac{3\pi r_h^8}{16(r_h^4-\alpha^2)^{3/2}}
		+
		\frac{3\pi^2T r_h^9}{8(r_h^4-\alpha^2)^{3/2}}
		\nonumber\\
		&
		-
		\frac{\pi^2}{2}r_h^3T\,
		{}_2F_1
		\left(
		-\frac{3}{4},
		\frac{3}{2};
		\frac{1}{4};
		\frac{\alpha^2}{r_h^4}
		\right).
		\label{eq:qt_on_shell_gibbs}
	\end{align}
	When the parametric curve is interpreted on shell, \(T\) is the Hawking
	temperature and \(P\) is determined by the equation of state
	\eqref{eq:qt_temperature_rh}. Below the critical temperature, the equation of state develops
	two spinodal points and the Gibbs free energy develops a swallowtail, as in the
	van der Waals small/large black hole transition, see Fig.~\ref{p1}.
	
	\subsection{Off-shell Gibbs free energy and entropy orientation}
	\label{subsec:qt_offshell_entropy_orientation}
	
	We now apply the general off-shell construction of Sec.~\ref{sec:offshell_general}. On a
	fixed-\((P,Q,\alpha)\) slice, define
	\begin{equation}
		G_{\rm off}^{\rm QT}(r_h;T_0,P,Q,\alpha)
		=
		M(r_h;P,Q,\alpha)
		-
		T_0S_W(r_h;\alpha),
		\label{eq:qt_offshell_gibbs}
	\end{equation}
	where \(T_0\) is the external bath temperature. Using the fixed-parameter first law
	\eqref{eq:qt_fixed_first_law}, one obtains
	\begin{equation}
		\frac{dG_{\rm off}^{\rm QT}}{dr_h}
		=
		\bigl[T(r_h)-T_0\bigr]S_W'(r_h).
		\label{eq:qt_goff_derivative}
	\end{equation}
	Therefore, as long as \(S_W'(r_h)\neq0\), stationary points of
	\(G_{\rm off}^{\rm QT}\) are precisely the black holes satisfying
	\begin{equation}
		T(r_h)=T_0 .
		\label{eq:qt_equilibrium_condition}
	\end{equation}
	
	The key question is the sign of \(S_W'(r_h)\). Starting from
	\eqref{eq:qt_wald_entropy}, let
	$z=\frac{\alpha^2}{r_h^4}$,
	then we have
	\begin{align}
		\frac{dS_W}{dr_h}
		=
		\frac{\pi^2}{2G}r_h^2
		\bigg[
		3\,
		{}_2F_1
		\left(
		\frac{3}{2},
		-\frac{3}{4};
		\frac{1}{4};
		z
		\right)
		-
		4z\frac{d}{dz}
		{}_2F_1
		\left(
		\frac{3}{2},
		-\frac{3}{4};
		\frac{1}{4};
		z
		\right)
		\bigg].
		\label{eq:qt_entropy_derivative_intermediate}
	\end{align}
	Using
	\begin{equation}
		\frac{d}{dz}
		{}_2F_1(a,b;c;z)
		=
		\frac{ab}{c}
		{}_2F_1(a+1,b+1;c+1;z),
		\label{eq:qt_hypergeometric_derivative_identity}
	\end{equation}
	together with the relevant contiguous relation, one finds
	\begin{align}
		&
		3\,
		{}_2F_1
		\left(
		\frac{3}{2},
		-\frac{3}{4};
		\frac{1}{4};
		z
		\right)
		-
		4z\frac{d}{dz}
		{}_2F_1
		\left(
		\frac{3}{2},
		-\frac{3}{4};
		\frac{1}{4};
		z
		\right)
		\nonumber\\
		&\hspace{2cm}
		=
		3\,
		{}_2F_1
		\left(
		\frac{3}{2},
		\frac{1}{4};
		\frac{1}{4};
		z
		\right).
		\label{eq:qt_contiguous_result}
	\end{align}
	Since
	\begin{equation}
		{}_2F_1(a,c;c;z)=(1-z)^{-a},
		\label{eq:qt_hypergeometric_simple_identity}
	\end{equation}
	we obtain the closed expression
	\begin{equation}
		S_W'(r_h)
		=
		\frac{3\pi^2r_h^8}
		{2G(r_h^4-\alpha^2)^{3/2}}.
		\label{eq:qt_entropy_derivative_closed}
	\end{equation}
	On the physical branch \(r_h>\sqrt{\alpha}\), this gives
	$S_W'(r_h)>0$.
	Thus the entropy coordinate is oriented in the same direction as \(r_h\)
	throughout the open physical branch. The endpoint \(r_h=\sqrt{\alpha}\) is singular because
	\(S_W'(r_h)\) diverges there, but the local thermodynamic analysis is performed at
	interior points of the branch.
	
	The off-shell Hessian at an equilibrium point \(r_h=r_i\) is
	\begin{equation}
		H_i^{\rm QT}
		=
		\left.
		\frac{d^2G_{\rm off}^{\rm QT}}{dr_h^2}
		\right|_{r_h=r_i}
		=
		\left.
		S_W'(r_h)T'(r_h)
		\right|_{r_h=r_i}.
		\label{eq:qt_hessian}
	\end{equation}
	Using \eqref{eq:qt_entropy_derivative_closed}, this becomes
	\begin{equation}
		H_i^{\rm QT}
		=
		\frac{3\pi^2r_i^8}
		{2G(r_i^4-\alpha^2)^{3/2}}
		T'(r_i).
		\label{eq:qt_hessian_explicit}
	\end{equation}
	Since the prefactor is strictly positive for \(r_i>\sqrt{\alpha}\), we have
	\begin{equation}
		{\rm sign}\,H_i^{\rm QT}
		=
		{\rm sign}\,T'(r_i).
		\label{eq:qt_sign_hessian_slope}
	\end{equation}
	Therefore, for this quasi-topological regular black hole on the physical branch, the generalized Hessian criterion reduces a posteriori to the usual temperature slope rule.
	
	The heat capacity on the same fixed-\((P,Q,\alpha)\) slice is
	\begin{equation}
		C_{P,Q,\alpha}
		=
		T
		\left(
		\frac{\partial S_W}{\partial T}
		\right)_{P,Q,\alpha}
		=
		T\frac{S_W'(r_h)}{T'(r_h)}.
		\label{eq:qt_heat_capacity}
	\end{equation}
	Away from spinodal points, where \(T'(r_h)\neq0\), and for \(T>0\) and,  
	\(S'_W(r_h)>0\), we obtain
	\begin{equation}
		{\rm sign}\,C_{P,Q,\alpha}
		=
		{\rm sign}\,T'(r_h)
		=
		{\rm sign}\,H_i^{\rm QT}.
		\label{eq:qt_heat_capacity_sign}
	\end{equation}
	A branch with positive temperature slope is locally stable, while a branch with
	negative temperature slope is locally unstable. This result agrees with the familiar
	Einstein gravity assignment \cite{Zhang:2025inl,Zhang:2026sht}, but the logic is different: the slope rule is
	derived from the positivity of the Wald entropy derivative on the chosen physical branch.
	
	\subsection{Branch structure and off-shell free energy profile}
	\label{subsec:qt_branch_structure}
	
	The equilibrium branches at fixed \((T_0,P,Q,\alpha)\) are the real roots of
	\begin{equation}
		T(r_h;P,Q,\alpha)-T_0=0
		\label{eq:qt_branch_roots}
	\end{equation}
	on the physical interval \(r_h>\sqrt{\alpha}\), with the additional requirement \(T_0>0\)
	for the canonical ensemble. The change from one to three real branches is controlled by
	the extrema of \(T(r_h)\), namely by the spinodal condition
	\begin{equation}
		T'(r_h;P,Q,\alpha)=0.
		\label{eq:qt_spinodal_condition}
	\end{equation}
	Since \eqref{eq:qt_temperature_rh} is linear in \(P\), the spinodal condition can be written as
	\begin{equation}
		P=P_{\rm sp}(r_h;q,\alpha).
		\label{eq:qt_spinodal_pressure}
	\end{equation}
	Moreover,
	\begin{equation}
		\frac{\partial T'}{\partial P}
		=
		\frac{4}{3}
		\frac{
			\sqrt{r_h^4-\alpha^2}\,
			(r_h^4+5\alpha^2)
		}{
			r_h^6
		}
		>0,
		\qquad
		r_h>\sqrt{\alpha}.
		\label{eq:qt_spinodal_pressure_derivative}
	\end{equation}
	For the parameter range considered below, one finds numerically that $P_{\rm sp}$ has a single maximum, which is identified with the critical point. Hence, for \(P<P_c\), there are two spinodal points and a finite interval of bath temperatures
	within which three equilibrium black holes coexist:
	\begin{equation}
		r_{h,s}<r_{h,m}<r_{h,l}.
		\label{eq:qt_three_branches}
	\end{equation}
	The outer branches have positive temperature slope and are locally stable, while the middle
	branch has negative temperature slope and is locally unstable:
	\begin{equation}
		{\rm sign}\,H_s^{\rm QT}=+,
		\qquad
		{\rm sign}\,H_m^{\rm QT}=-,
		\qquad
		{\rm sign}\,H_l^{\rm QT}=+.
		\label{eq:qt_branch_stability_ordering}
	\end{equation}
	
\begin{figure}[htbp]
		\centering
		\includegraphics[width=0.6\linewidth]{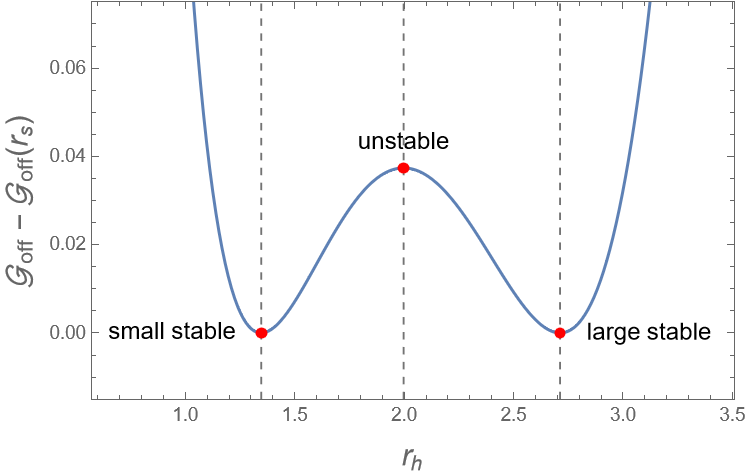}
		\caption{
			Off-shell Gibbs free energy at \(P=0.8P_c\) for \(q=2/3\) and \(\alpha=9/25\), shifted by the small black hole minimum. The bath temperature is chosen as \(T_0=T_{\rm coex}\simeq0.125587\). The three stationary points are the roots of \(T(r_h)=T_0\). The small and large branches are two local minima with equal free energy,
			while the intermediate branch is a local maximum.
		}
		\label{p2}
	\end{figure}

	For illustration, take
	$q=\frac{2}{3}$,
	and $\alpha=\frac{9}{25}$.
	The critical point is obtained from
	\begin{equation}
		\left(\frac{\partial P}{\partial v}\right)_T=0,
		\qquad
		\left(\frac{\partial^2P}{\partial v^2}\right)_T=0.
		\label{eq:qt_critical_conditions}
	\end{equation}
	Numerically, we have
	$v_c\simeq 2.425463,
	r_{h,c}\simeq 1.819097$,
	and
	$T_c\simeq 0.137859,
	P_c\simeq 0.0234737$.
	At
	$P=0.8P_c$,
	the coexistence temperature is
	$T_0=T_{\rm coex}\simeq 0.125587$.
	The stationary points of \(G_{\rm off}^{\rm QT}\) are located at
	\begin{equation}
		r_{h,s}\simeq 1.346955,
		\qquad
		r_{h,m}\simeq 1.997594,
		\qquad
		r_{h,l}\simeq 2.712503.
		\label{eq:qt_stationary_points_example}
	\end{equation}
	The small and large black holes are local minima of \(G_{\rm off}^{\rm QT}\), while the
	intermediate black hole is a local maximum. At coexistence, the two minima have equal
	off-shell Gibbs free energy, see Fig.~\ref{p2}. This provides a direct visualization of the Hessian criterion:
	local minima have \(H_i^{\rm QT}>0\), while the local maximum has \(H_i^{\rm QT}<0\).
	
	\subsection{Critical cusp from the off-shell free energy}
	\label{subsec:qt_cusp}
	
	The off-shell free energy also gives a compact description of the local critical behavior.
	Near the critical point, the entropy itself is the natural coordinate on the one-dimensional
	configuration space. We define
	\begin{equation}
		x=S_W(r_h)-S_W(r_{h,c}),
		\qquad
		p=P-P_c,
		\qquad
		\theta=T_0-T_c.
		\label{eq:qt_entropy_coordinate}
	\end{equation}
	Because \(S_W'(r_{h,c})>0\), the coordinate transformation between \(r_h\) and \(x\) is
	regular near the critical point. By defining 
	\begin{equation}
		F(x;p,\theta)
		=
		G_{\rm off}^{\rm QT}(r_h(x);T_0,P,Q,\alpha),
		\label{eq:qt_F_definition}
	\end{equation}
	we have 
	\begin{equation}
		\frac{\partial F}{\partial x}
		=
		T(x,P)-T_0.
		\label{eq:qt_F_x}
	\end{equation}
	
	For a nondegenerate mean-field critical point, one finds
	\begin{equation}
		T_x|_c=0,
		\qquad
		T_{xx}|_c=0,
		\qquad
		T_{xxx}|_c\neq0,
		\label{eq:qt_entropy_coordinate_critical_conditions}
	\end{equation}
	and the pressure direction is nondegenerate if
	\begin{equation}
		T_{xP}|_c\neq0,
		\qquad
		T_P|_c\neq0.
		\label{eq:qt_pressure_nondegenerate_conditions}
	\end{equation}
	Expanding \eqref{eq:qt_F_x} gives
	\begin{equation}
		\frac{\partial F}{\partial x}
		=
		h
		+
		Apx
		+
		Bx^3
		+
		\cdots,
		\label{eq:qt_cusp_stationary_expansion}
	\end{equation}
	where
	\begin{equation}
		h=T_P|_c\,p-\theta+\cdots,
		\qquad
		A=T_{xP}|_c=\frac{T_{rP}|_c}{S'_W(r_{h,c})},
		\qquad
		B=\frac{1}{6}T_{xxx}|_c=\frac{T_{rrr}|_c}{6 [S'_W(r_{h,c})]^3}.
		\label{eq:qt_cusp_coefficients}
	\end{equation}
	Integrating once in \(x\), the singular part of the off-shell free energy becomes
	\begin{equation}
		F_{\rm sing}
		=
		hx
		+
		\frac{A}{2}px^2
		+
		\frac{B}{4}x^4
		+
		\cdots .
		\label{eq:qt_cusp_free_energy}
	\end{equation}
	This is the local \(A_3\) cusp normal form.
	
	For the numerical parameters \eqref{eq:qt_stationary_points_example}, one finds
	$T_r|_c=0,
	T_{rr}|_c=0$,
	and
	$T_{rrr}|_c\simeq 0.108464, 
	T_{rP}|_c\simeq 1.40385,
	T_P|_c\simeq 2.38253$.
	Since
	$S_W'(r_{h,c})\simeq 49.8722>0$,
	the entropy coordinate coefficients are
	$A\simeq 0.0281490,
	B\simeq 1.4573\times10^{-7}$.
	
	Along the coexistence curve, the two stable extrema have equal off-shell free energy.
	In the cusp normal form, this selects the field free direction
	$h=0$
	at leading order. The stationary equation then becomes
	\begin{equation}
		x(Ap+Bx^2)=0.
		\label{eq:qt_cusp_stationary_equation}
	\end{equation}
	For \(A>0\) and \(B>0\), the two nonzero extrema exist for \(p<0\) and satisfy
	\begin{equation}
		x_{\pm}
		=
		\pm
		\left(
		-\frac{Ap}{B}
		\right)^{1/2}
		+
		O(p).
		\label{eq:qt_x_pm}
	\end{equation}
	Since
	\begin{equation}
		r_h-r_{h,c}
		=
		\frac{x}{S_W'(r_{h,c})}
		+
		O(x^2),
		\label{eq:qt_rh_x_relation}
	\end{equation}
	the small/large branch separation obeys
	\begin{equation}
		r_{h,l}-r_{h,s}
		\propto
		|p|^{1/2}.
		\label{eq:qt_branch_separation_p}
	\end{equation}
	Moreover, along \(h=0\), we have 
	\begin{equation}
		\theta=T_P|_c\,p+O(p^2).
		\label{eq:qt_theta_p_relation}
	\end{equation}
	Thus, in terms of the reduced temperature
	\begin{equation}
		t=1-\frac{T_0}{T_c},
		\label{eq:qt_reduced_temperature}
	\end{equation}
	one obtains
	\begin{equation}
		r_{h,l}-r_{h,s}
		\propto
		t^{1/2}.
		\label{eq:qt_branch_separation_t}
	\end{equation}
	The mean-field branch-separation exponent \(1/2\) along the coexistence curve is therefore a direct consequence of the off-shell cusp structure. More generally, let \(\mathcal{O}(r_h)\) be any
	branch observable that is smooth near the critical point and satisfies
	\begin{equation}
		\mathcal{O}_l-\mathcal{O}_s
		=
		\mathcal{O}'(r_{h,c})(r_{h,l}-r_{h,s})+\cdots
		\propto t^{1/2}.
	\end{equation}
	The Lyapunov exponent of unstable null orbits provides such an observable
	when the photon-sphere equation is nondegenerate at the critical point.
	Under these assumptions, the Lyapunov branch difference inherits the
	thermodynamic \(1/2\) exponent \cite{Xie:2025auj}. It is therefore not an independent critical
	exponent, but a consequence of the same small/large branch splitting that
	controls the off-shell free energy. This provides a direct bridge between
	black hole phase transitions and geometry based diagnostics such as
	photon sphere observables \cite{Zhang:2019glo,Xu:2019yub,Li:2019dai,Kumara:2019xgd,Du:2022cgw,Kumar:2024dws,Yang:2023hci,Lyu:2023axv,Kumara:2024wge,Du:2024fqu,Shukla:2024vhl,Gogoi:2024wyq,Chen:2025nbx,Karthik:2025jrv,Awal:2025qmu,Kumar:2025cos,Guo:2025qin,Bezboruah:2025modmax,Ali:2025pfdm,Su:2024gxp,Becar:2026ned,Zhang:2025cdx}.
	
	\section{Lovelock black holes}
	\label{sec:lovelock_black_holes}
	
	The quasi-topological regular black hole discussed in Sec.~\ref{sec:qt_black_hole}
	provides an example in which the Wald entropy is nontrivial but remains monotonic on the
	physical branch. As a result, the off-shell Hessian criterion reduces a posteriori to the
	ordinary temperature slope rule. In this section we show that this reduction is not generic
	in higher-curvature gravity.
	
	Lovelock black holes provide a particularly transparent setting for this purpose. Their
	thermodynamics is known in closed form, and the derivative of the Wald entropy can be
	written directly in terms of the Lovelock polynomial. This makes explicit how the entropy
	measure on the one-dimensional configuration space can change orientation. When this
	happens, the usual Einstein gravity temperature slope stability rule is no longer reliable;
	it must be replaced by the off-shell Hessian criterion.
	
	\subsection{Lovelock action and static black hole branches}
	\label{subsec:lovelock_action_metric}
	
	We consider Lovelock gravity in \(D\) spacetime dimensions. The action is
	\begin{equation}
		I_{\rm Lov}
		=
		\frac{1}{16\pi G}
		\int d^Dx\sqrt{-g}
		\sum_{n=0}^{K} c_n {\cal L}_n ,
		\label{eq:lovelock_action}
	\end{equation}
	where \(K\le\lfloor(D-1)/2\rfloor\), \(c_n\) are Lovelock couplings, and purely topological terms are not included.  In addition, the term 
	\begin{equation}
		{\cal L}_n
		=
		\frac{1}{2^n}
		\delta^{\mu_1\nu_1\cdots\mu_n\nu_n}_{\rho_1\sigma_1\cdots\rho_n\sigma_n}
		R^{\rho_1\sigma_1}{}_{\mu_1\nu_1}
		\cdots
		R^{\rho_n\sigma_n}{}_{\mu_n\nu_n}
		\label{eq:lovelock_density}
	\end{equation} 
	is the \(n\)-th Lovelock density \cite{Lovelock:1971yv}. In particular, \({\cal L}_0=1\), \({\cal L}_1=R\), and
	\({\cal L}_2\) is the Gauss--Bonnet density.
	
	It is convenient to use the standard dimensionally rescaled Lovelock couplings \(\hat c_n\), normalized by \(\hat c_1=1\), and to define the Lovelock polynomial \cite{Jacobson:1993xs,Cai:2003kt,Kastor:2010gq}. The constant term \(\hat c_0\) is the rescaled cosmological coupling, while for \(n\ge2\) the \(\hat c_n\) include the conventional dimension-dependent Lovelock factors
	\begin{equation}
		\Upsilon(g)
		=
		\sum_{n=0}^{K}\hat c_n g^n .
		\label{eq:lovelock_polynomial}
	\end{equation}
	The cosmological constant is included in the constant term. For AdS black holes we write
	\begin{equation}
		\hat c_0
		=
		\frac{1}{l^2}
		=
		\frac{16\pi G P}{(D-1)(D-2)},
		\label{eq:lovelock_pressure}
	\end{equation}
	where \(P\) is the thermodynamic pressure.
	
	Here we  take the static topological black hole ansatz
	\begin{equation}
		ds^2
		=
		-f(r)dt^2
		+
		\frac{dr^2}{f(r)}
		+
		r^2 d\Sigma_{D-2,k}^2,
		\qquad
		k=1,0,-1 ,
		\label{eq:lovelock_metric}
	\end{equation}
	where \(d\Sigma_{D-2,k}^2\) is the line element on a \((D-2)\)-dimensional constant-curvature
	space with volume \(\Sigma_{D-2}\). We define
	\begin{equation}
		g(r)=\frac{k-f(r)}{r^2}.
		\label{eq:lovelock_g_definition}
	\end{equation}
	For this ansatz, the Lovelock field equations reduce to the algebraic relation
	\begin{equation}
		\Upsilon(g(r))
		=
		\frac{\mu}{r^{D-1}},
		\label{eq:lovelock_field_equation}
	\end{equation}
	where \(\mu\) is proportional to the black hole mass.
	
	Different roots of \eqref{eq:lovelock_field_equation} correspond to different Lovelock
	branches. A physical branch must be selected before any thermodynamic statement is made.
	In particular, the branch should be real outside the horizon, should approach the chosen
	AdS vacuum at infinity, and should satisfy the relevant branch regularity and ghost-free
	conditions. The local off-shell analysis below applies once such a branch has been chosen.
	
	The horizon radius \(r_h\) is determined by
	\begin{equation}
		f(r_h)=0.
		\label{eq:lovelock_horizon_condition}
	\end{equation}
	Therefore, one can obtain 
	\begin{equation}
		g_h
		\equiv
		g(r_h)
		=
		\frac{k}{r_h^2}.
		\label{eq:lovelock_gh}
	\end{equation}
	Substituting this into \eqref{eq:lovelock_field_equation}, we obtain 
	\begin{equation}
		\mu
		=
		r_h^{D-1}
		\Upsilon\!\left(\frac{k}{r_h^2}\right).
		\label{eq:lovelock_mu_horizon}
	\end{equation}
	The ADM mass, interpreted as enthalpy in the extended phase space, is then
	\begin{equation}
		M(r_h)
		=
		\frac{(D-2)\Sigma_{D-2}}{16\pi G}
		r_h^{D-1}
		\Upsilon\!\left(\frac{k}{r_h^2}\right).
		\label{eq:lovelock_mass}
	\end{equation}
	
	\subsection{Thermodynamics and Wald entropy}
	\label{subsec:lovelock_thermodynamics}
	
	The Hawking temperature is obtained from the surface gravity,
	\begin{equation}
		T
		=
		\frac{f'(r_h)}{4\pi}.
		\label{eq:lovelock_temperature_definition}
	\end{equation}
	Using \(f(r)=k-r^2g(r)\) and differentiating the algebraic equation
	\eqref{eq:lovelock_field_equation}, one finds
	\begin{equation}
		T(r_h)
		=
		\frac{r_h}{4\pi}
		\left[
		\frac{(D-1)\Upsilon(g_h)}{\Upsilon'(g_h)}
		-
		2g_h
		\right],
		\qquad
		g_h=\frac{k}{r_h^2}.
		\label{eq:lovelock_temperature}
	\end{equation}
	This expression assumes
	\begin{equation}
		\Upsilon'(g_h)\neq0.
		\label{eq:lovelock_upsilon_prime_nonzero}
	\end{equation}
	Points where \(\Upsilon'(g_h)=0\) are branch degeneracy points and require separate
	treatment.
	
	The entropy of Lovelock black holes is the Wald/Jacobson--Myers entropy,
	\begin{equation}
		S_W(r_h)
		=
		\frac{\Sigma_{D-2}r_h^{D-2}}{4G}
		\sum_{n=1}^{K}
		\frac{n(D-2)}{D-2n}
		\hat c_n
		\left(
		\frac{k}{r_h^2}
		\right)^{n-1}.
		\label{eq:lovelock_entropy}
	\end{equation}
	Equivalently,
	\begin{equation}
		S_W(r_h)
		=
		\frac{\Sigma_{D-2}(D-2)}{4G}
		\sum_{n=1}^{K}
		\frac{n\hat c_n k^{\,n-1}}{D-2n}
		r_h^{D-2n}.
		\label{eq:lovelock_entropy_equiv}
	\end{equation}
	Differentiating with respect to \(r_h\), we obtain
	\begin{equation}
		S_W'(r_h)
		=
		\frac{\Sigma_{D-2}(D-2)}{4G}
		\sum_{n=1}^{K}
		n\hat c_n k^{\,n-1}
		r_h^{D-2n-1}.
		\label{eq:lovelock_entropy_derivative_sum}
	\end{equation}
	Using the Lovelock polynomial, this derivative can be written in the compact form
	\begin{equation}
		S_W'(r_h)
		=
		\frac{\Sigma_{D-2}(D-2)}{4G}
		r_h^{D-3}
		\Upsilon'(g_h).
		\label{eq:lovelock_entropy_derivative}
	\end{equation}
	This equation is central to the discussion below. It shows that the entropy orientation along the black hole branch is controlled by \(\Upsilon'(g_h)\). In Einstein gravity,
	\(\Upsilon'(g_h)=1\), and the entropy is positively oriented for all \(r_h>0\). In Lovelock
	gravity, however, for planar horizons, \(g_h=0\) and \(\Upsilon'(0)=\hat c_1=1\), so the entropy orientation is necessarily positive. Possible reversals of \(S_W'(r_h)\) are therefore tied to non-planar horizons. Moreover, a change of sign along a continuous branch would require crossing \(\Upsilon'(g_h)=0\), where the branch becomes degenerate. Hence the sign of \(S_W'(r_h)\) should be discussed branch segment by branch segment, after imposing the appropriate regularity and ghost-free conditions.
	
	The fixed-coupling first law takes the form
	\begin{equation}
		dM
		=
		T\,dS_W
		+
		V\,dP ,
		\label{eq:lovelock_first_law_fixed_couplings}
	\end{equation}
	where the Lovelock couplings other than the cosmological constant are held fixed. If the
	higher-curvature couplings are also varied, the first law is enlarged to
	\begin{equation}
		dM
		=
		T\,dS_W
		+
		V\,dP
		+
		\sum_{n\geq2}\Psi_n\,d\hat c_n .
		\label{eq:lovelock_first_law_full}
	\end{equation}
	In this work we study local branch stability on fixed-\((P,\hat c_n)\) slices. Thus the
	relevant one-dimensional first law is
	\begin{equation}
		dM=T\,dS_W.
		\label{eq:lovelock_fixed_slice_first_law}
	\end{equation}
	
	For completeness, we verify this relation explicitly. Differentiating the mass
	\eqref{eq:lovelock_mass} at fixed pressure and fixed Lovelock couplings gives
	\begin{equation}
		\frac{dM}{dr_h}
		=
		\frac{(D-2)\Sigma_{D-2}}{16\pi G}
		r_h^{D-2}
		\left[
		(D-1)\Upsilon(g_h)
		-
		2g_h\Upsilon'(g_h)
		\right].
		\label{eq:lovelock_mass_derivative}
	\end{equation}
	Combining \eqref{eq:lovelock_mass_derivative} with
	\eqref{eq:lovelock_entropy_derivative}, one finds
	\begin{equation}
		\frac{dM/dr_h}{dS_W/dr_h}
		=
		\frac{r_h}{4\pi}
		\left[
		\frac{(D-1)\Upsilon(g_h)}{\Upsilon'(g_h)}
		-
		2g_h
		\right]
		=
		T(r_h),
		\label{eq:lovelock_first_law_check}
	\end{equation}
	which is precisely the fixed-parameter first law.
	
	The thermodynamic volume follows from differentiating \(M\) with respect to \(P\) at fixed
	\(r_h\) and fixed higher-curvature couplings:
	\begin{equation}
		V
		=
		\left(
		\frac{\partial M}{\partial P}
		\right)_{r_h,\hat c_{n\geq1}}
		=
		\frac{\Sigma_{D-2}r_h^{D-1}}{D-1}.
		\label{eq:lovelock_volume}
	\end{equation}
	
	\subsection{Off-shell Gibbs free energy}
	\label{subsec:lovelock_offshell}
	
	We now construct the off-shell Gibbs free energy in direct analogy with
	Sec.~\ref{sec:offshell_general} and Sec.~\ref{sec:qt_black_hole}. At fixed pressure and
	fixed Lovelock couplings, we  define
	\begin{equation}
		G_{\rm off}^{\rm Lov}(r_h;T_0,P,\hat c_n)
		=
		M(r_h;P,\hat c_n)
		-
		T_0S_W(r_h;\hat c_n),
		\label{eq:lovelock_offshell_gibbs}
	\end{equation}
	where \(T_0\) is the external bath temperature. This function assigns a free energy value
	to each horizon configuration \(r_h\) on the chosen Lovelock branch, regardless of whether
	the configuration is in equilibrium with the bath.
	
	Using the fixed-parameter first law \eqref{eq:lovelock_fixed_slice_first_law}, we obtain
	\begin{equation}
		\frac{dG_{\rm off}^{\rm Lov}}{dr_h}
		=
		\frac{dM}{dr_h}
		-
		T_0\frac{dS_W}{dr_h}
		=
		\left[
		T(r_h)-T_0
		\right]
		S_W'(r_h).
		\label{eq:lovelock_goff_derivative}
	\end{equation}
	Therefore, whenever
	\begin{equation}
		S_W'(r_h)\neq0,
		\label{eq:lovelock_nonzero_entropy_derivative}
	\end{equation}
	the stationary condition
	$
	\frac{dG_{\rm off}^{\rm Lov}}{dr_h}=0$
	is equivalent to the thermal equilibrium condition
	\begin{equation}
		T(r_h)=T_0.
		\label{eq:lovelock_equilibrium_condition}
	\end{equation}
	Thus the usual temperature equation survives, but it is derived from the off-shell
	variational principle constructed with the Wald entropy.
	
	Equation \eqref{eq:lovelock_goff_derivative} also shows why entropy turning points must
	be treated carefully. If
	\begin{equation}
		S_W'(r_h)=0,
		\label{eq:lovelock_entropy_turning_point}
	\end{equation}
	then \(dG_{\rm off}^{\rm Lov}/dr_h=0\) independently of whether \(T(r_h)=T_0\). Such a
	point is not, by itself, a thermal equilibrium black hole. Rather, it is a degeneracy point
	of the entropy coordinate on the chosen branch. In Lovelock gravity, by
	\eqref{eq:lovelock_entropy_derivative}, this condition is equivalent to
	\begin{equation}
		\Upsilon'(g_h)=0.
		\label{eq:lovelock_upsilon_prime_zero}
	\end{equation}
	These points should be excluded from the ordinary one-dimensional stability analysis or
	treated as branch boundaries unless an independent analysis shows that the branch can be
	continued through them.
	
	\subsection{Hessian stability criterion}
	\label{subsec:lovelock_hessian}
	
	The local stability of an equilibrium Lovelock black hole is determined by the second
	variation of \(G_{\rm off}^{\rm Lov}\). Differentiating
	\eqref{eq:lovelock_goff_derivative} once more gives
	\begin{equation}
		\frac{d^2G_{\rm off}^{\rm Lov}}{dr_h^2}
		=
		T'(r_h)S_W'(r_h)
		+
		\left[
		T(r_h)-T_0
		\right]
		S_W''(r_h).
		\label{eq:lovelock_second_derivative}
	\end{equation}
	At an equilibrium point \(r_h=r_i\), where \(T(r_i)=T_0\), the second term vanishes.
	Therefore the local Hessian is
	\begin{equation}
		H_i^{\rm Lov}
		\equiv
		\left.
		\frac{d^2G_{\rm off}^{\rm Lov}}{dr_h^2}
		\right|_{r_h=r_i}
		=
		\left.
		S_W'(r_h)T'(r_h)
		\right|_{r_h=r_i}.
		\label{eq:lovelock_hessian}
	\end{equation}
	The equilibrium branch is locally stable against horizon radius fluctuations on the fixed
	\((P,\hat c_n)\) slice if
	\begin{equation}
		H_i^{\rm Lov}>0,
		\label{eq:lovelock_stability_positive}
	\end{equation}
	and locally unstable if
	\begin{equation}
		H_i^{\rm Lov}<0.
		\label{eq:lovelock_stability_negative}
	\end{equation}
	
	Using \eqref{eq:lovelock_entropy_derivative}, the Hessian can be written as
	\begin{equation}
		H_i^{\rm Lov}
		=
		\frac{\Sigma_{D-2}(D-2)}{4G}
		r_i^{D-3}
		\Upsilon'\!\left(\frac{k}{r_i^2}\right)
		T'(r_i).
		\label{eq:lovelock_hessian_upsilon}
	\end{equation}
	This expression makes the central point transparent. The local stability of a Lovelock
	black hole is not controlled by the temperature slope alone. It is controlled by the product
	of the temperature slope and the entropy derivative.
	
	The same conclusion follows from the heat capacity. On a fixed-\((P,\hat c_n)\) slice, we have 
	\begin{equation}
		C_{P,\hat c_n}
		=
		T
		\left(
		\frac{\partial S_W}{\partial T}
		\right)_{P,\hat c_n}
		=
		T
		\frac{S_W'(r_h)}{T'(r_h)}.
		\label{eq:lovelock_heat_capacity}
	\end{equation}
	For positive temperature, we obtain 
	\begin{equation}
		{\rm sign}\,C_{P,\hat c_n}
		=
		{\rm sign}
		\left[
		S_W'(r_h)T'(r_h)
		\right]
		=
		{\rm sign}\,H_i^{\rm Lov}.
		\label{eq:lovelock_heat_capacity_hessian}
	\end{equation}
	Thus the heat-capacity criterion and the off-shell Hessian criterion are equivalent in the
	positive temperature canonical sector. The advantage of the off-shell formulation is that it
	shows directly why the entropy derivative is part of the stability data.
	
	In Einstein gravity, one finds that 
	\begin{equation}
		\Upsilon(g)=\hat c_0+g,
		\qquad
		\Upsilon'(g)=1.
		\label{eq:lovelock_einstein_polynomial}
	\end{equation}
	Hence
	\begin{equation}
		S_W'(r_h)>0
		\label{eq:lovelock_einstein_entropy_positive}
	\end{equation}
	on the physical branch, and
	\begin{equation}
		{\rm sign}\,H_i
		=
		{\rm sign}\,T'(r_i).
		\label{eq:lovelock_einstein_slope_rule}
	\end{equation}
	This is the familiar temperature slope rule.
	
	In Lovelock gravity, however, the sign of \(\Upsilon'(g_h)\) is model dependent and
	branch dependent. The temperature slope rule is valid only in regions where
	\begin{equation}
		S_W'(r_h)>0
		\quad
		\Longleftrightarrow
		\quad
		\Upsilon'(g_h)>0.
		\label{eq:lovelock_positive_entropy_orientation}
	\end{equation}
	If instead
	$S_W'(r_h)<0$,
	the usual temperature slope stability assignment is reversed:
	\begin{equation}
		{\rm sign}\,H_i^{\rm Lov}
		=
		-
		{\rm sign}\,T'(r_i),
		\qquad
		T'(r_i)\neq0.
		\label{eq:lovelock_slope_reversal_general}
	\end{equation}
	A branch with positive temperature slope is then locally unstable, while a branch with
	negative temperature slope is locally stable. This is the precise sense in which the
	Einstein gravity slope rule is not a fundamental stability criterion. It is a consequence of
	a positively oriented entropy measure.

	The planar case provides a useful check. For \(k=0\), all terms with \(k^{n-1}\) vanish in
	\eqref{eq:lovelock_entropy_derivative_sum} except the Einstein term \(n=1\). The entropy
	therefore reduces to the area law
	\begin{equation}
		S_W(r_h)
		=
		\frac{\Sigma_{D-2}r_h^{D-2}}{4G},
		\label{eq:planar_lovelock_entropy}
	\end{equation}
	\begin{equation}
		S_W'(r_h)
		=
		\frac{\Sigma_{D-2}(D-2)}{4G}
		r_h^{D-3}
		>0.
		\label{eq:planar_lovelock_entropy_derivative}
	\end{equation}
	Therefore, we can see that 
	\begin{equation}
		{\rm sign}\,H_i^{\rm Lov}
		=
		{\rm sign}\,T'(r_i)
		\label{eq:planar_lovelock_slope_rule}
	\end{equation}
	for planar Lovelock black holes.
	
	This does not mean that higher-curvature terms have no effect on planar black hole
	thermodynamics. They can still modify the temperature function, the equation of state,
	and the global phase structure through the Lovelock polynomial. However, they do not
	reverse the entropy orientation on the one-dimensional horizon radius branch. Consequently,
	the ordinary temperature slope rule remains valid in the planar case.
	
	\subsection{Entropy orientation and branch admissibility}
	
	The discussion above shows that a reversal of the usual temperature-slope
	criterion would require
	$S'_W(r_h)<0$.
	For Lovelock black holes with constant-curvature horizons, this condition is
	equivalent to
	\begin{equation}
		\Upsilon'(g_h)<0,\qquad g_h=\frac{k}{r_h^2}.
	\end{equation}
	This is because
	\begin{equation}
		S'_W(r_h)
		=
		\frac{\Sigma_{D-2}(D-2)}{4G}
		r_h^{D-3}\Upsilon'(g_h).
	\end{equation}
	This observation also shows that the entropy orientation is not an arbitrary
	thermodynamic feature: it is tied directly to the Lovelock branch structure.
	
	Let the black hole exterior interpolate between the horizon value \(g_h\) and
	an asymptotic AdS vacuum \(\Lambda\), where
	\begin{equation}
		\Upsilon(g_\infty)=0,
	\end{equation}
	a ghost-free AdS vacuum requires \cite{Boulware:1972yco}
	\begin{equation}
		\Upsilon'(g_\infty)>0.
	\end{equation}
	If one nevertheless had \(S'_W(r_h)<0\), then we have 
	\begin{equation}
		\Upsilon'(g_h)<0.
	\end{equation}
	
	By continuity, there would exist a point \(g_\ast\) between \(g_h\) and
	\(\Lambda\) such that
	$
	\Upsilon'(g_\ast)=0.
	$
	Differentiating the Lovelock equation
	\begin{equation}
		\Upsilon(g(r))=\frac{\mu}{r^{D-1}}
	\end{equation}
	gives
	\begin{equation}
		\Upsilon'(g)g'(r)
		=
		-\frac{(D-1)\mu}{r^D}.
	\end{equation}
	Thus, at a zero of \(\Upsilon'(g)\), the radial derivative of the branch
	generically diverges. Such a point is the usual Lovelock branch singularity.
	If it lies outside the horizon, the corresponding black hole branch is not a
	regular exterior solution. Therefore a ghost-free and branch-regular Lovelock
	black hole exterior cannot have
	$
	S'_W(r_h)<0.
	$
	Equivalently,
	\[
	\text{ghost-free and branch-regular exterior}
	\quad\Longrightarrow\quad
	\Upsilon'(g_h)>0
	\quad\Longrightarrow\quad
	S'_W(r_h)>0.
	\]
	
	This gives a useful refinement of the Hessian criterion. In a purely formal
	Lovelock branch with \(S'_W<0\), the off-shell Hessian would indeed reverse
	the temperature-slope assignment,
	$
	\operatorname{sign}H_i=-\operatorname{sign}T'(r_i).
	$
	However, in the physically admissible Lovelock branches satisfying the usual
	ghost-free and branch-regularity requirements, the entropy orientation is
	positive and the ordinary temperature-slope rule is protected. Hence, in
	constant-curvature Lovelock black holes, a would-be reversal of the slope rule
	is better interpreted as a diagnostic of a pathological or inadmissible branch
	rather than as a generic physical phase.
	
	For illustration, here we consider the cubic Lovelock polynomial
	\begin{equation}
		\Upsilon(g)=\frac{1}{10}+g+g^2+\frac15 g^3
	\end{equation}
	in \(D=7\) with hyperbolic horizon topology \(k=-1\). At \(r_h=1\), one has
	\(g_h=-1\), and hence
	\begin{equation}
		\Upsilon(g_h)=-\frac{1}{10},\qquad
		\Upsilon'(g_h)=-\frac25<0.
	\end{equation}
	Then we can see that the  Wald entropy is nevertheless positive,
	\begin{equation}
		S_W(1)
		=
		\frac{\Sigma_5}{4G}
		\left(1-\frac{10}{3}+3\right)
		=
		\frac{\Sigma_5}{6G}>0,
	\end{equation}
	and the temperature is also positive,
	\begin{equation}
		T(1)
		=
		\frac{1}{4\pi}
		\left[
		6\frac{\Upsilon(g_h)}{\Upsilon'(g_h)}
		-2g_h
		\right]
		=
		\frac{7}{8\pi}>0.
	\end{equation}
	Moreover \(T'(1)=35/(8\pi)>0\), while \(S'_W(1)<0\), so the off-shell Hessian
	is negative even though the temperature slope is positive. This example
	demonstrates the algebraic possibility of the entropy-orientation reversal.
	However, it should not be regarded as a fully admissible black hole branch:
	the ghost-free AdS root has positive \(\Upsilon'\), whereas the horizon has
	negative \(\Upsilon'\), implying a zero of \(\Upsilon'\) and hence a branch
	singularity between the horizon and infinity. Alternatively, choosing a
	negative-slope AdS vacuum avoids the sign crossing but gives a Boulware--Deser
	ghost branch. Thus the example is useful as a diagnostic test of the Hessian criterion, but it
	also shows why the reversed-orientation segment is excluded once the standard
	Lovelock branch-admissibility conditions are imposed.

	\section{Conclusions}
	\label{sec:conclusions}
	
	We have studied the local thermodynamic stability of higher-curvature black
	holes using the off-shell Gibbs free energy built from the Wald entropy. On a
	fixed-parameter branch, the stationary condition of
	\(G_{\rm off}=M-T_0S_W\) gives the equilibrium equation \(T(r_h)=T_0\) whenever
	\(S'_W(r_h)\neq0\). The second variation gives
	$H_i=S'_W(r_i)T'(r_i)$. A useful way to summarize the result is that the ordinary temperature slope rule is not fundamental but protected. It is automatic in Einstein gravity because the area entropy is positively oriented, whereas in higher-curvature gravity it survives only when the Wald entropy remains positively oriented on the physical branch. Thus, whenever the familiar slope assignment is recovered, its validity already encodes nontrivial information about the entropy measure and branch admissibility.
	
	We applied this criterion to two examples. For the five-dimensional charged
	regular AdS black hole in quasi-topological gravity, the Wald entropy is
	nontrivial but monotonic on the physical branch, so \(S'_W>0\) and the ordinary
	temperature slope rule is recovered. For Lovelock black holes,
	\(S'_W\propto\Upsilon'(g_h)\), and this factor may be negative on certain non-planar branch segments. In regions with \(S'_W<0\), the stability assignment is reversed:
	positive temperature slope gives local instability, while negative temperature
	slope gives local stability. This reversal should be understood as a structural consequence of the off-shell Hessian criterion; in concrete Lovelock solutions, the physical admissibility of a negative orientation segment must still be checked against positive temperature, nonnegative entropy, branch regularity, and ghost-free conditions.
	
	Our analysis is local and branch-dependent. It assumes a fixed-parameter first law,
	a well defined entropy functional, and a chosen physical branch. Within
	these limits, the off-shell Hessian criterion provides a compact way to decide when
	the Einstein gravity temperature slope rule survives and when the more general
	condition \(H_i=S'_W(r_i)T'(r_i)\) is required. More importantly, it turns the search
	for a genuinely physical reversal of the slope rule into a sharply formulated
	problem: one must find an admissible branch with \(T>0\), nonnegative Wald entropy,
	and \(S'_W<0\), while still satisfying regularity and ghost-free conditions. The
	Lovelock analysis shows that such a reversal is excluded on branch-regular and
	ghost-free Lovelock exteriors, but the criterion leaves open the possibility that
	more general higher-curvature theories, hairy black holes, Einstein-scalar-tensor black holes, or more general horizon geometries \cite{Celik:2026ngw,Hajian:2023bhq,Martinez:2005di,Liu:2026kvm} may realize a physically
	allowed negative entropy orientation.

	\acknowledgments
	
	We are grateful to Shi-Hao Zhang for valuable discussions. This work was supported by the National Natural Science Foundation of China under Grants No. 12475051, No. 12375051, and No. 12421005; the science and technology innovation Program of Hunan Province under grant No. 2024RC1050; the Natural Science Fund of Hunan Province under Grant No. 2026JJ20019; and the innovative research group of Hunan Province under Grant No. 2024JJ1006.
	
	\appendix
	
	\section{Additional high-curvature examples}
	\label{app}
	
	\subsection{Einstein--Gauss--Bonnet AdS black holes}
	
	As a first standard check, consider Einstein--Gauss--Bonnet gravity in
	\(D\geq5\). In the Lovelock-polynomial convention used in Sec.~\ref{sec:lovelock_black_holes}:
	$\Upsilon(g)=\hat c_0+g+\alpha_{\rm GB}g^2$ .
	For a topological black hole with horizon curvature \(k=1,0,-1\), the Wald
	entropy is
	\[
	S_W^{\rm GB}
	=
	\frac{\Sigma_{D-2}r_h^{D-2}}{4G}
	\left[
	1+\frac{2(D-2)\alpha_{\rm GB}k}{(D-4)r_h^2}
	\right].
	\]
	Hence
	\[
	\frac{dS_W^{\rm GB}}{dr_h}
	=
	\frac{\Sigma_{D-2}(D-2)}{4G}
	r_h^{D-5}
	\left(r_h^2+2\alpha_{\rm GB}k\right).
	\]
	The off-shell Hessian criterion therefore gives
	\[
	H_i^{\rm GB}
	=
	\frac{\Sigma_{D-2}(D-2)}{4G}
	r_i^{D-5}
	\left(r_i^2+2\alpha_{\rm GB}k\right)T'(r_i).
	\]
	
	For \(k=1\) and \(\alpha_{\rm GB}>0\), the entropy orientation is positive for
	all \(r_h>0\), and the ordinary temperature-slope rule is recovered. For
	\(k=0\), the Gauss--Bonnet correction drops out of the Wald entropy and
	\(S'_W>0\) again. The only possible negative-orientation region for
	\(\alpha_{\rm GB}>0\) occurs for hyperbolic horizons:
	$
	k=-1, r_h^2<2\alpha_{\rm GB}.
	$
	However, the entropy itself is then
	\[
	S_W^{\rm GB}
	=
	\frac{\Sigma_{D-2}r_h^{D-2}}{4G}
	\left[
	1-\frac{2(D-2)\alpha_{\rm GB}}{(D-4)r_h^2}\right],
	\]
	so the condition \(S_W\geq0\) requires
	\[
	r_h^2\geq \frac{2(D-2)}{D-4}\alpha_{\rm GB}.
	\]
	Since
	\[
	\frac{2(D-2)}{D-4}\alpha_{\rm GB}>2\alpha_{\rm GB}
	\qquad (D>4),
	\]
	the formal region \(S'_W<0\) lies inside the negative-entropy region. Thus the
	Einstein--Gauss--Bonnet example gives a useful consistency check: a would-be
	reversal of the temperature-slope criterion is present algebraically, but it is
	excluded once nonnegative entropy is imposed.
	
	\subsection{Constant-curvature \(f(R)\) AdS black holes}
	
	As a second check, we consider an \(f(R)\) theory admitting a constant-curvature
	black hole with
	\[
	R=R_0,\qquad f_R(R_0)\equiv \left.\frac{df}{dR}\right|_{R_0}=\text{constant}.
	\]
	For such solutions, the Wald entropy is simply a constant rescaling of the area
	entropy,
	\[
	S_W^{f(R)}
	=
	\frac{f_R(R_0)}{4G}\,\Sigma_{D-2}r_h^{D-2}.
	\]
	Therefore, we have 
	\[
	\frac{dS_W^{f(R)}}{dr_h}
	=
	\frac{f_R(R_0)}{4G}(D-2)\Sigma_{D-2}r_h^{D-3}.
	\]
	The corresponding off-shell Hessian is
	\[
	H_i^{f(R)}
	=
	\frac{f_R(R_0)}{4G}(D-2)\Sigma_{D-2}r_i^{D-3}T'(r_i).
	\]
	
	Thus the sign of the entropy orientation is controlled by \(f_R(R_0)\). If
	\(f_R(R_0)>0\), the effective Newton constant
	$
	G_{\rm eff}=\frac{G}{f_R(R_0)}
	$
	is positive, the Wald entropy is positively oriented, and the usual
	temperature-slope rule is recovered:
	$
	\operatorname{sign}H_i^{f(R)}=\operatorname{sign}T'(r_i).
	$
	If \(f_R(R_0)<0\), the slope assignment would be formally reversed. However,
	this would also correspond to a negative effective Newton constant and a
	negative Wald entropy for positive horizon area. Therefore, in the physically
	admissible constant-curvature \(f(R)\) sector with \(f_R(R_0)>0\), the
	Einstein-gravity temperature-slope rule is again protected.
	
	This example shows that the off-shell Hessian criterion is not tied to the
	Lovelock polynomial. What matters is the orientation of the Wald entropy along
	the thermodynamic branch. In \(f(R)\) gravity this orientation is controlled by
	\(f_R(R_0)\), while in Lovelock gravity it is controlled by \(\Upsilon'(g_h)\).

\end{document}